\documentstyle[12pt,psfig]{article}
\begin{document}

\title{\bf Energy Levels of Classical Interacting Fields in a Finite Domain in 
$1 + 1$ Dimension}

\author{ J. A. Espich\'an Carrillo$^{\S}$\footnote{E-mail address: espichan@ifi.unicamp.br} \ and  \ A. Maia Jr.$^{\P}$\footnote{E-mail address: maia@ime.unicamp.br}\\[5mm]
$^{\S}$Instituto de F\'{\i}sica ``Gleb Wathagin'',  UNICAMP 13.081-970 - Campinas -\\ S.P. - Brazil.\\
$^{\P}$Instituto de Matem\'atica, UNICAMP 13.081-970 - Campinas - S.P. - Brazil.}

\maketitle

\noindent
{\bf Abstract.} \ We study the behavior of bound energy levels for the case of two classical interacting fields $\phi$ and $\chi$ in a finite domain (box) in $(1 + 1)$ dimension on which we impose Dirichlet boundary conditions (DBC). The total Lagrangian contain a $\frac{\lambda}{4}\phi^4$ self-interaction and an interaction term given by $g\phi^2\chi^2$. We calculate the energy eigenfunctions and its correspondent eigenvalues and study their dependence on the size of the box $(L)$ as well on the free parameters of the Lagrangian: mass ratio $\beta = \frac{M^{2}_{\chi}}{M^{2}_{\phi}}$, and interaction coupling constants $\lambda$ and $g$. We show that for some configurations of the above parameters, there exists critical sizes of the box for which instability points of the field $\chi$ appear.

\newpage

\baselineskip 0.65cm

\noindent
{\large \bf 1. Introduction}

\vspace*{0.5cm}

It is well known that quantum physical systems can alter significantly their behavior when placed inside cavities. A modern paradigm is the famous Casimir Effect [1] and more recently the so called Cavity Quantum Electrodynamics [2].

From a mathematical point of view part of these studies can be translated into the general setting of differential equations for quantum fields, on which are imposed suitable boundary conditions, in order to know wavefunctions and energy eigenvalues. The above mentioned subjects are essentially of quantum nature. Nevertheless it is well known that, for several applications, quantum fields can be thought as classical fields on which are added quantum corrections [3]. In this sense, although at a classical level, we can get a lot of information about the system under study.

In this work we study the influence of boundary conditions on bound energy levels of a classical system of fields described by Lagrangian density 
\begin{eqnarray}
{\cal L} =  \frac{1}{2}(\partial_{\mu}\phi)^2 + 
\frac{1}{2}M_{\phi}^2\phi^2 - \frac{\lambda}{4}\phi^4 +
\frac{1}{2}(\partial_{\mu}\chi)^2 + \frac{1}{2}M_{\chi}^2\chi^2 
- g\phi^2\,\chi^2,
\label{p}
\end{eqnarray}
where $\lambda$, $g$ are coupling constants. 

Before to continue, we want to make an important observation. In (\ref{p}), the Lagrangian density of the field $\chi$, i.e. ${\cal L} = \frac{1}{2}(\partial_{\mu}\chi)^2 \ + \ \frac{1}{2}M_{\chi}^2\chi^2$, \ does not have a state of least energy since its associated Hamiltonian is not positive definite. This, clearly, is due to the ``wrong'' sign of the mass term. One way of to solve this would be by adding a term of self-interaction for the field $\chi$, as usually happens in theories with spontaneous symmetry breaking. Of course we could also keep the original positive sign in the mass term. This will lead to different results from this work, but all tecniques are all the same. In this work, we chose to keep the positive sign of the mass term and verify when the interaction term of the Lagrangian, given by $- g\phi^2\,\chi^2$, leads to a lower bound to the Hamiltonian density. See Appendix for details. 

We consider only the simpler case of fields $\phi$ and $\chi$ inside a finite box (interval) in $(1 + 1)$ dimension. Of course all discussion can be generalized to higher dimensions with a number of new differentials equations and boundary conditions related to geometry of the box. A quantum version of the theory, in a semiclassical approach will be done elsewhere.

The Equation of motion (E.O.M.) for the two fields are given by
\begin{eqnarray}
-\partial^{\mu}\partial_{\mu}\chi + M_{\chi}^{2}\chi - 2\,g\phi^{2}\chi = 0,
\label{sn1}
\end{eqnarray}
\begin{eqnarray}
-\partial^{\mu}\partial_{\mu}\phi + M_{\phi}^{2}\phi - \lambda\phi^{3}= 0.
\label{sn2}
\end{eqnarray}

In Eq.(3) we have neglected the term $2\,g\phi\chi^2$ which can be interpreted as the back-reaction of field $\chi$ on the mass term of $\phi$. This can be achieved if we impose for example that $|\chi|<< \frac{M_{\phi}}{\sqrt{2\,g}}$. Of course other regimes can be studied from Eq. (\ref{sn2}) by adopting different approximations.

In a previous work [4] we have studied, in $(1 + 1)$ dimension, the case in which the field $\phi(x)$, unlike $\chi(x)$, is not influenced by finite boundary conditions of the box. So $\phi(x)$ is the Kink solution of Eq. (\ref{sn2}) in ($-\infty , +\infty$) [5]. In this case we can think $\chi$-field is placed in a fixed potential $\phi^{2}$ and we showed that a level splitting appears (bifurcation point). This could be interpreted, in a semi-classical version of the theory, as $\chi$-particle creation induced by squeezing the box below a critical size.

In this work we take into account the same boundary conditions for both fields $\phi$ and $\chi$. This means that the potential $\phi^{2}$ depends on the size of the box. So we study the behavior of the energy levels of $\chi$ field by running the parameters $L, \beta , \lambda , g$, where $\beta = \frac{M^{2}_{\chi}}{M^{2}_{\phi}}$ is the mass ratio and $L$ is the box size. Of course, the box size is an external parameter of the theory. We show below that classical instabilities appear for a critical size of box.

A family of static solutions of the classsical equation of motion to the field $\phi(x)$ are given by ${\rm sn}$-type elliptic functions [6]
\begin{eqnarray}
\phi_c(x) \ = \ \pm\frac{M_{\phi}\sqrt{2c}}{\sqrt{\lambda}\sqrt
{1+\sqrt{1-2c}}}\,
{\rm sn}\left(\frac{M_{\phi}x}{\sqrt{2}}\sqrt{1+\sqrt{1-2c}},l\right),
\label{sn3}
\end{eqnarray} 
where $c$ is a parameter belonging to interval $(0,\frac{1}{2}]$ and
\begin{eqnarray}
l = \frac{1}{-1+\frac{1+\sqrt{1-2c}}{c}}. 
\label{sn4}
\end{eqnarray}

Clearly, $l \in (0,1]$. There exist another family of solutions. See reference [6]
for details. The above one was chosen because we can impose Dirichlet boundary conditions on their solutions. In general $\phi_{c}$ is a function of $x-x_0$, but we can put $x_0 = 0$ without loss of generality.

So, the equation of motion for the field $\chi$ can be written as
\begin{eqnarray*}
\left( -\partial^{\mu}\partial_{\mu} + M^2_{\chi} - \frac{4\,g\,M^2_{\phi} c}
{\lambda(1+\sqrt{1-2c})}{\rm sn}^{2}\left(\frac{M_{\phi}x}{\sqrt{2}}
\sqrt{1+\sqrt{1-2c}},l\right)\right)\chi = 0.
\end{eqnarray*}

Since we are interested in stationary solutions we can write
$\chi(x,t) = e^{-i\omega\,t}\psi(x)$, where $\omega$ are energy eigenvalues. With this, the previous equation we can be writen in the form
\begin{eqnarray}
\frac{d^2}{dx^{2}}\psi(x) + \left( M^2_{\chi} +\omega^2 - \frac{4\,g\,M^2_{\phi} c}{\lambda(1+\sqrt{1-2c})}
{\rm sn}^{2}\left(\frac{M_{\phi}x}{\sqrt{2}}
\sqrt{1+\sqrt{1-2c}},l\right)\right)\psi(x) = 0.
\label{sn5}
\end{eqnarray}

In the next section we calculate the energy eigenvalues of Eq. (\ref{sn5}) as well study their dependence on the parameters of the theory, namely, $g$, $\lambda$, $\beta = \frac{M^{2}_{\chi}}{M^{2}_{\phi}}$, $l$. For short 
$\omega \equiv \omega(\lambda,g,\beta;l)$, where the semicolon means $l$ is an external parameter of the theory. In section 5 we study the level shifts induced by changing the box size and interpret the results. In section 6 we conclude with some comments and list some topics for future work.

\vspace*{0.9cm}

\noindent
{\large \bf 2. Lam\'e Equation and Boundary Conditions} 

\vspace*{0.5cm}

In this section we study the bound levels of two interacting fields which are confined inside a box (interval in our case) of length $L$. We impose Dirichlet boundary conditions (DBC) for both fields $\phi$ and $\chi$ (or $\psi$).

In [6] the authors showed that imposing DBC on the field $\phi$, given by (\ref{sn3}), confined to a box of size $L$, it must satisfy the condition
\begin{eqnarray}
M_{\phi}L = 4\sqrt{1+l}\,K(l), 
\label{sn6}
\end{eqnarray}
where $4K(l)$ is a period of the Jacobi Elliptic Functions ${\rm sn}(u,l)$ [7]. To get the above equation we impose (DBC) at $x_1 = -\frac{L}{2}$ and $x_2 = \frac{L}{2}$. Observe that since the solutions (\ref{sn3})
are odd functions, the points $x_0 =0$ does not give any information and it is sufficient to impose DBC just, say, at $x_2 = \frac{L}{2}$.

Since we take the same boundary conditions for both fields (same kind of confinement), this implies that the same $l = l(L)$ obtained from the above equation must be substituted in the $\chi$-field boundary conditions, in order to find its energy eigenvalues.

Of course different boundary conditions can be imposed independently on the fields $\phi$ and $\chi$. For example, in our previous work [4] we have studied the extreme case that the box boundaries are transparent for the field $\phi$, while $\chi$ satisfies DBC.

We start making the changes of variables, 
\begin{eqnarray*}
\alpha = \frac{M_{\phi}x}{\sqrt{2}}\sqrt{1+\sqrt{1-2c}} \ \ \ \ \ \ \ \ 
\mbox{and} 
\ \ \ \ \ \ \ \ \omega^2 = \frac{(E - 2)}{2}M_{\phi}^2,
\end{eqnarray*}
in equation (\ref{sn5}) which can then be rewritten as
\begin{eqnarray}
\frac{d^2}{d \alpha^2}\psi(\alpha) \ = \ \left( \frac{8\,g\,c}
{\lambda(1+\sqrt{1-2c})^2}{\rm sn}^{2}(\alpha,l) - 2\frac{(M_{\chi}^2 - M_{\phi}^2)}{M_{\phi}^2 \sqrt{1+\sqrt{1-2c}}} - \frac{E}{(1+\sqrt{1-2c})}\right)\psi(\alpha).
\label{sn7}
\end{eqnarray}

On the other hand, from (\ref{sn4}) we have
\begin{eqnarray}
c=\frac{2l}{(l+1)^2}.
\label{sn8}
\end{eqnarray}

Thus (\ref{sn7}) reduces to
\begin{eqnarray}
\frac{d^2}{d \alpha^2}\psi(\alpha) \ = \ \left( 4\frac{g}{\lambda}\,l\,
{\rm sn}^{2}(\alpha,l) - \frac{(M_{\chi}^2 - M_{\phi}^2)}{M_{\phi}^2}(1+l)  - 
\frac{E(1+l)}{2}\right)\psi(\alpha).
\label{sn9}
\end{eqnarray}

This differential equation has some important special properties. Since $g$ and $\lambda$ are positive we can write, without lost of generality 
\begin{eqnarray}
4\frac{g}{\lambda} \equiv n(n+1),
\label{sn10}
\end{eqnarray}
where $n$ is a positive real number.

So, the above equation can be rewritten as
\begin{eqnarray}
\frac{d^2}{d \alpha^2}\psi(\alpha) \ = \ \left( n(n+1)l\,{\rm sn}^{2}(\alpha,l) - (\beta - 1)(1+l)  - \frac{E(1+l)}{2}\right)\psi(\alpha),
\label{sn11}
\end{eqnarray}
where we have defined the adimensional mass ratio parameter as $\beta \equiv \frac{M^{2}_{\chi}}{M^{2}_{\phi}}$.

This is a Generalized Lam\'e differential equation. In the literature the general form of this type of equation is given by [8]\footnote{In our notation we take the parameter of the Jacobian Elliptic Function as $k$ (with $k > 0$) instead of $k^2$ as in reference [8].}
\begin{eqnarray}
\frac{d^2}{d \alpha^2}\Lambda(\alpha) \ = \ \left( n(n+1)\,k\,{\rm sn}^{2}(\alpha,k) + C \right)\Lambda(\alpha),
\label{sn12}
\end{eqnarray}
where $n$ is a positive real number, $k$ is the parameter of the Jacobian 
Elliptic Function ${\rm sn}$, and $C$ is an arbitrary constant. 

It is well known that Lam\'e differential equation and more generally the Hill's equation presents stability as well instability bands in the plane of parameters $(k,C)$ in the notation of equation (\ref{sn12}). This stability is related to the spatial dependence of the solution. On the other hand, we are interested in stability for the time dependence. Using Floquet's theory the solution can be written as
\begin{eqnarray*}
\chi(x,t) = e^{-i\omega\,t}e^{irx}p(x),
\end{eqnarray*}
where $p(x)$ is a periodic function. In our approach below (section 3) we show that even in the case for eigenvalues of Lam\'e equation describing stable solutions for the spatial part $(r^2 > 0)$, we can have unstable solutions in time, that is $\omega^2 < 0$.

Comparing (\ref{sn11}) with (\ref{sn12}), we obtain
\begin{eqnarray*}
C = - (\beta - 1)(1+l) - \frac{E(1+l)}{2} \ \ \ \ \ \ \ \mbox{and} \ \ \ \ \ \ \ 
k = l.
\end{eqnarray*}

For the purpose we have in  mind we only consider in this work the case that $n$ is a positive integer. The case with $n$ real, although more interesting, leads to eigenvalues which are difficult to calcule exactly and a full numerical treatment is necessary in order to obtain the eigenvalues we are interested to. For $n$ integer the equation (\ref{sn11}) or (\ref{sn12}) is called simply Lam\'e differential equation and can be solved analitically. So, our results can be used also as a test for the numerical solutions of more realistic cases which do not have exact solutions.

For $n$ a positive integer the general solution of Eq.(\ref{sn11}) is given by 
\begin{eqnarray*}
\psi(\alpha) = A\,E^{m}_{n}(\alpha)  +  B\,F^{m}_{n}(\alpha),
\end{eqnarray*}
where $A$ and $B$ are arbitrary constants and $E^{m}_{n}(\alpha)$ and  $F^{m}_{n}(\alpha)$ are Lam\'e Functions of the first and second kind respectively [8]. The parameter $m$ ranges on $\{ -n,-n+1,...,n-1,n \}$.
Moreover, when $n$ is a  positive integer, if one of the solutions of the Lam\'e equation is a polynomial, then the second solution must be an infinite series. The polynomial solution is given by $E^{m}_{n}(\alpha)$ and the series solution by $F^{m}_{n}(\alpha)$ [8].

In this work, we restrict our study only to polynomial solutions. In other words, we search for solutions whose growth at infinite, be polynomial. So our solutions are given only by
\begin{eqnarray}
\psi(\alpha) = A\,E^{m}_{n}(\alpha).
\label{sn14}
\end{eqnarray}
 
Since we are interested only in eigenvalues, in the following we drop the arbitrary constant from the eigenfunctions (\ref{sn14}).

Below we show the first eigenfunctions $(n = 1,2,3)$ of Eq. (\ref{sn11}) which are given by Lam\'e Functions. The case for $n$ continuous will not considered in this work. Observe that the case $n = 0$, in principle, could be considered. But from (\ref{sn10}), $n = 0$ implies $g = 0$  which in turn leads to a Hamiltonian which is not positive definite. Therefore the case $n = 0$ will be discarded. Of course by Eq. (\ref{sn10}) there is a minimal strengh for the coupling constant namely, $g = \frac{\lambda}{2}$ $(\mbox{for} \ n = 1)$.  On the other hand, strong coupling $(n \rightarrow\infty)$ leads to a new and more complicated Lam\'e Functions. In this work we analyse only the case for $n$ small, that is, the coupling constant has a moderate strengh. We will see that, yet for these few cases a rich phenomenology for the bound energy levels emerges. 

\vspace*{0.9cm}

\noindent
{\large \bf 3. Eigenfunctions and Eigenvalues of $\psi$-Field}

\vspace*{0.5cm}

Using the results from reference [8], for $E^{m}_{n}(\alpha)$, as well as, the form defined for the eigenvalues $H$, namely
\begin{eqnarray*}
H \ = \ \frac{1}{l}C, 
\end{eqnarray*}
we list below the eigenfunctions $\psi(\alpha)$ (\ref{sn14}) and their eigenvalues for  Eq.(\ref{sn11}), i.e. 

\vspace*{0.5cm}

{\bf CASE I: $n = 1$ $(g = \frac{\lambda}{2})$}
\begin{eqnarray*}
1) \ \ \ \psi_1(x,l)  = \,{\rm sn}(\frac{M_{\phi}x}{\sqrt{1+l}},l),
\end{eqnarray*}
\begin{eqnarray} 
H_{1}^{-1}(l) = -1 - \frac{1}{l}.
\label{sn16}
\end{eqnarray}
\begin{eqnarray*}
2) \ \ \ \psi_2(x,l)  = \,\sqrt{{\rm sn}^{2}(\frac{M_{\phi}x}{\sqrt{1+l}},l)-1},
\end{eqnarray*}
\begin{eqnarray} 
H_{1}^{0}(l) = - \frac{1}{l}.
\label{sn17}
\end{eqnarray}
\begin{eqnarray*}
3) \ \ \ \psi_3(x,l)  = \,\sqrt{\frac{l\,{\rm sn}^2(\frac{M_{\phi}x}{\sqrt{1+l}},l) - 1}{l}}, 
\end{eqnarray*}
\begin{eqnarray} 
H_{1}^{1}(l) = - 1.
\label{sn18}
\end{eqnarray}

\vspace*{0.5cm}

{\bf CASE II: $n = 2$ $(g = \frac{3}{2}\lambda)$}
\begin{eqnarray*}
1) \ \ \ \psi_1(x,l)  = \,{\rm sn}(\frac{M_{\phi}x}{\sqrt{1+l}},l)\sqrt{{\rm sn}^2(\frac{M_{\phi}x}{\sqrt{1+l}},l)-1}, 
\end{eqnarray*}
\begin{eqnarray} 
H_{2}^{-1}(l) = -\frac{1}{l}(4 + l).
\label{sn19}
\end{eqnarray}
\begin{eqnarray*}
2) \ \ \ \psi_2(x,l) = \,{\rm sn}(\frac{M_{\phi}x}{\sqrt{1+l}},l)\sqrt{\frac{l\,{\rm sn}^2(\frac{M_{\phi}x}{\sqrt{1+l}},l) - 1}{l}}, 
\end{eqnarray*}
\begin{eqnarray}
H_{2}^{0}(l) = -\frac{1}{l}(4l + 1).
\label{sn20}
\end{eqnarray}
\begin{eqnarray*}
3) \ \ \ \psi_3(x,l) = \,\sqrt{{\rm sn}^2(\frac{M_{\phi}x}{\sqrt{1+l}},l) -1}\,\sqrt{\frac{l\,{\rm sn}^2(\frac{M_{\phi}x}{\sqrt{1+l}},l) - 1}{l}}, 
\end{eqnarray*}
\begin{eqnarray} 
H_{2}^{1}(l) = -\frac{1}{l}(l + 1).
\label{sn21}
\end{eqnarray}
\begin{eqnarray*}
4) \ \ \ \psi_4(x,l) = \,{\rm sn}^2(\frac{M_{\phi}x}{\sqrt{1+l}},l) - \frac{1}{1 + l + \sqrt{l^2 - l + 1}}, 
\end{eqnarray*}
\begin{eqnarray} 
H_{2}^{2}(l) = -\frac{2}{l}(1 + l + \sqrt{l^2 - l + 1}).
\label{sn22}
\end{eqnarray}
\begin{eqnarray*}
5) \ \ \ \psi_5(x,l) = \, {\rm sn}^2(\frac{M_{\phi}x}{\sqrt{1+l}},l) -\frac{1}{1 + l - \sqrt{l^2 - l + 1}}, 
\end{eqnarray*}
\begin{eqnarray} 
H_{2}^{-2}(l) = -\frac{2}{l}(1 + l - \sqrt{l^2 - l + 1}).
\label{sn23}
\end{eqnarray}

\vspace*{0.5cm}

{\bf CASE III: $n = 3$ $(g = 3\lambda)$}
\begin{eqnarray*}
1) \ \ \ \psi_1(x,l) = \,\sqrt{{\rm sn}^2(\frac{M_{\phi}x}{\sqrt{1+l}},l)-1}
\left( {\rm sn}^2(\frac{M_{\phi}x}{\sqrt{1+l}},l) + \frac{1}{-\sqrt{{l}^{2} -l + 4} -l - 2}\right),
\end{eqnarray*}
\begin{eqnarray} 
H_{3}^{-2} = -\frac{2l + 5}{l} - \frac{2}{l}\sqrt{l^2 - l + 4}.
\label{sn24}
\end{eqnarray}
\begin{eqnarray*}
2) \ \ \ \psi_2(x,l) = \,\sqrt{ \frac{l\,{\rm sn}^2(\frac{M_{\phi}x}{\sqrt{1+l}},l)- 1}{l}}\left( {\rm sn}^2(\frac{M_{\phi}x}{\sqrt{1+l}},l) + \frac{1}{-\sqrt{4{l}^{2} -l + 1} -2l - 1}\right),
\end{eqnarray*}
\begin{eqnarray} 
H_{3}^{-1} = -\frac{5l + 2}{l} - \frac{2}{l}\sqrt{4l^2 - l + 1}.
\label{sn25}
\end{eqnarray}
\begin{eqnarray*}
3) \ \ \ \psi_3(x,l)  = \,{\rm sn}(\frac{M_{\phi}x}{\sqrt{1+l}},l)\sqrt{{\rm sn}^2(\frac{M_{\phi}x}{\sqrt{1+l}},l)-1}\,\sqrt{\frac{l\,{\rm sn}^2(\frac{M_{\phi}x}{\sqrt{1+l}},l) - 1}{l}}, 
\end{eqnarray*}
\begin{eqnarray} 
H_{3}^{0}(l) = -\frac{4}{l}(1 + l).
\label{sn26}
\end{eqnarray}
\begin{eqnarray*}
4) \ \ \ \psi_4(x,l) = \,{\rm sn}(\frac{M_{\phi}x}{\sqrt{1+l}},l)\left( {\rm sn}^2(\frac{M_{\phi}x}{\sqrt{1+l}},l) + \frac{3}{\sqrt{4(l - 1)^{2} + l} 
-2l -2} \right)
\end{eqnarray*}
\begin{eqnarray}
H_{3}^{1} = -\frac{5}{l}(l + 1) + \frac{2}{l}\sqrt{4(l - 1)^{2} + l}.
\label{sn27}
\end{eqnarray}
\begin{eqnarray*}
5) \ \ \ \psi_5(x,l) = \,\sqrt{{\rm sn}^2(\frac{M_{\phi}x}{\sqrt{1+l}},l)-1}
\left( {\rm sn}^2(\frac{M_{\phi}x}{\sqrt{1+l}},l) + \frac{1}{\sqrt{{l}^{2} -l + 4} -l - 2}\right),
\end{eqnarray*}
\begin{eqnarray} 
H_{3}^{2} = -\frac{2l + 5}{l} + \frac{2}{l}\sqrt{l^2 - l + 4}.
\label{sn28}
\end{eqnarray}
\begin{eqnarray*}
6) \ \ \ \psi_6(x,l) = \,\sqrt{ \frac{l\,{\rm sn}^2(\frac{M_{\phi}x}{\sqrt{1+l}},l) - 1}{l}}\left( {\rm sn}^2(\frac{M_{\phi}x}{\sqrt{1+l}},l) + \frac{1}{\sqrt{4{l}^{2} -l + 1} -2l - 1}\right),
\end{eqnarray*}
\begin{eqnarray} 
H_{3}^{3} = -\frac{5l + 2}{l} + \frac{2}{l}\sqrt{4l^2 - l + 1}.
\label{sn29}
\end{eqnarray}
\begin{eqnarray*}
7) \ \ \ \psi_7(x,l) = \,{\rm sn}(\frac{M_{\phi}x}{\sqrt{1+l}},l)\left( {\rm sn}^2(\frac{M_{\phi}x}{\sqrt{1+l}},l) + \frac{3}{-\sqrt{4(l - 1)^{2} + l} 
-2l -2}\right)
\end{eqnarray*}
\begin{eqnarray}
H_{3}^{-3} = -\frac{5}{l}(l + 1) - \frac{2}{l}\sqrt{4(l - 1)^{2} + l}.
\label{sn30}
\end{eqnarray}

Where, in all cases we substituted with help of (\ref{sn8}), 
$\alpha = \frac{M_{\phi}x}{\sqrt{1+l}}$.

For $l\in(0,1]$ all the above eigenvalues $H_{n}^{m}$ are negative.

\vspace*{0.9cm}

\noindent
{\large \bf 4. Eigenvalues for Dirichlet Boundary Conditions}

\vspace*{0.5cm}

In this section we obtain the energy eigenvalues $\omega^2$ by 
imposing DBC at $x=\pm\frac{L}{2}$ on the solutions $\psi_{s}$ above. Thus we obtain relations as $l \equiv l(L)$. We have made use of the relation $\omega^2 =\frac{(E-2)}{2}M^2_{\phi}$. 

An important observation is now in order.
In [6] was determined a minimum value for the relation $M_{\phi}L$, when $l \rightarrow 0$, namely, $2\pi$. Using (\ref{sn6}) (the same as  (\ref{sn35}) below), this result can be checked easily. Likewise observe from (\ref{sn8}) that if $l \rightarrow 0$ then $c \rightarrow 0$ and then by (\ref{sn3}) we have that $\phi \rightarrow 0$. 
So, below the minimum value $M_{\phi}L = 2\pi$ the field $\phi$ vanishes. 
Therefore, in our calculation the only consistent eigenvalues are those that satisfy the condition
$M_{\phi}L \geq 2\pi$, that is, those for which the field $\phi$ does not vanish. This argument will be used in several cases below. We now turn to the calculation of the energy eigenvalues for the eigenfunctions $\psi_{s} $.

\vspace*{0.5cm}

{\bf CASE I: $n = 1$ $(g = \frac{\lambda}{2})$}
\begin{eqnarray}
1) \ \ \ \omega^{2}_{1}(\beta) = (1 - \beta)M^{2}_{\phi}.
\label{sn31}
\end{eqnarray}

Observe that in this case $\omega_{1}^{2}$ does not depend on $l$.
\begin{eqnarray*}
2) \ \ \ \omega^{2}_{2}(l,\beta) = (\frac{1}{1 + l} - \beta)M^{2}_{\phi},
\end{eqnarray*}
where $l$ satisfies
\begin{eqnarray*}
M_{\phi}L = 2\sqrt{1 + l}K(l).
\end{eqnarray*}

Note that the $l = l(L)$ solution of this equation is not the same to that one from (\ref{sn6}). As previously mentioned, we are interested to obtain energy eigenvalues of field $\chi$ with the same $l = l(L)$ used for field $\phi$. Thus $\omega_{2}^{2}$ must be discarded for the case $n =1$.
\begin{eqnarray}
3) \ \ \ \omega^{2}_{3}(l,\beta) = (\frac{l}{1 + l} - \beta)M^{2}_{\phi},
\label{sn32}
\end{eqnarray}
where $l$ satisfies
\begin{eqnarray}
{\rm sn}^{2}(\frac{M_{\phi}L}{2\sqrt{1+l}},l)  = \frac{1}{l}.
\label{sn33}
\end{eqnarray}

Since ${\rm sn}^{2}(\frac{M_{\phi}L}{2\sqrt{l+1}},l) \leq 1$, $l$ should satisfies $l \geq 1$. Since $l \in (0,1]$, only $l = 1$ $(L = \infty)$ is solution of the Eq. (\ref{sn33}).

\vspace*{0.5cm}

{\bf CASE II: $n = 2$ $(g = \frac{3}{2}\lambda)$}
\begin{eqnarray}
1) \ \ \ \omega_{1}^{2}(l,\beta) = (\frac{4 + l}{1 + l} - \beta)M^2_{\phi},
\label{sn34}
\end{eqnarray}
where $l$ satisfies
\begin{eqnarray}
M_{\phi}L = 4\sqrt{1 + l}K(l)
\label{sn35}
\end{eqnarray}
or
\begin{eqnarray}
M_{\phi}L = 2\sqrt{1 + l}K(l).
\label{sn36}
\end{eqnarray}

Notice, that in this case only the Eq. (\ref{sn35}) satisfies the condition that the same $l = l(L)$ must be used for both fields $\phi \ \mbox{and} \ \chi$. According to this, solutions of equations like (\ref{sn36}) must be discarded. So $\omega^{2}_{1}$ is an allowed eigenvalue with $l$ given by (\ref{sn35}).

\begin{eqnarray}
2) \ \ \ \omega_{2}^{2}(l,\beta) = (\frac{4\,l + 1}{1 + l} - \beta)M^2_{\phi},
\label{sn37}
\end{eqnarray}
with $l$ satisfies (\ref{sn35}) or also 
\begin{eqnarray}
{\rm sn}^2(\frac{M_{\phi}L}{2\sqrt{l + 1}},l) = \frac{1}{l}.
\label{sn38}
\end{eqnarray}

Here as in the case of the Eq. (\ref{sn33}), only $l = 1$ $(L = \infty)$ is solution of this equation. So $\omega^{2}_{2}$ is also an allowed eigenvalue with $l$ given by (\ref{sn35}).
\begin{eqnarray}
3) \ \ \ \omega_{3}^{2}(\beta) = (1 - \beta)M_{\phi}^{2},
\label{sn39}
\end{eqnarray}
observe that $\omega_{3}^{2}$ is independent of $l$ and therefore of $L$.
\begin{eqnarray}
4) \ \ \ \omega_{4}^{2}(l,\beta) = (\frac{(2 - \beta)(1 + l) + 2\sqrt{l^2 - l + 1}}{1 + l})M^2_{\phi},
\label{sn40}
\end{eqnarray}
where $l$ satisfies   
\begin{eqnarray}
{\rm sn}^2(\frac{M_{\phi}L}{2\sqrt{l+1}},l) = \frac{1}{1 + l +
\sqrt{l^2 - l + 1}}.
\label{sn41}
\end{eqnarray}

A numerical analysis shows that for $l \in (0,1]$ we obtain, from the above equation that the values of $M_{\phi}L$ belongs to the interval $(1.57,1.86)$. So, the eigenvalue $\omega_{4}^{2}$ is not a consistent solution since for this case $M_{\phi}L < 2\pi$.
\begin{eqnarray}
5) \ \ \ \omega_{5}^{2}(l,\beta) = (\frac{(2 -\beta)(1 + l) - 2\sqrt{l^2 - l + 1}}
{1 + l})M^2_{\phi},
\label{sn42}
\end{eqnarray}
where $l$ satisfies  
\begin{eqnarray}
{\rm sn}^2(\frac{M_{\phi}L}{2\sqrt{l+1}},l) = \frac{1}{1 + l -
\sqrt{l^2 - l + 1}}.
\label{sn43}
\end{eqnarray}

This relation has only the solution, $l = 1$ ($L = \infty$).

\vspace*{0.8cm}

{\bf CASE III: $n = 3$ $(g = 3\lambda)$}

\begin{eqnarray}
1) \ \ \ \omega_{1}^{2}(l,\beta) = (\frac{2l + 5 + 2\sqrt{l^{2} - l + 4}}{1 + l} - \beta)M^2_{\phi}.
\label{sn44}
\end{eqnarray}

Since we have a product in (\ref{sn24}) we get or $l$ satisfying Eq. (\ref{sn36}), which should be discarded, or $l$ satisfies
\begin{eqnarray}
{\rm sn}^2(\frac{M_{\phi}L}{2\sqrt{l+1}},l) = \frac{1}{l + 2 +
\sqrt{l^2 - l + 4}}.
\label{sn45}
\end{eqnarray}

A numerical analysis shows that for $l\in (0,1]$ $M_{\phi}L$ belongs to the interval $(1.04,1.36)$ and then $M_{\phi}L < 2\pi$. Therefore the eigenvalue $\omega_{6}^{2}$ is not a consistent solution.
\begin{eqnarray}
2) \ \ \ \omega_{2}^{2}(l,\beta) = (\frac{5l + 2 + 2\sqrt{4\,l^{2} - l + 1}}{1 + l} - \beta)M^2_{\phi},
\label{sn46}
\end{eqnarray}
with $l$ satisfying 
\begin{eqnarray}
{\rm sn}^2(\frac{M_{\phi}L}{2\sqrt{l + 1}},l) = \frac{1}{l} \ \ \ \ \ \ \ \
\mbox{or} \ \ \ \ \ \ \ \ {\rm sn}^2(\frac{M_{\phi}L}{2\sqrt{l+1}},l) = \frac{1}{2l + 1 + \sqrt{4l^2 - l + 1}}.
\label{sn47}
\end{eqnarray}

Observe that the first equation is satisfied only for $l = 1$ ($L = \infty $), and for the second equation is possible to show numerically that $M_{\phi}L$ belongs to the interval $(1.36,1.57)$. Thus this equation will not be considered since $M_{\phi}L < 2\pi$.
\begin{eqnarray}
3) \ \ \ \omega_{3}^{2}(l,\beta) = (4  - \beta)M^2_{\phi}.
\label{sn48}
\end{eqnarray}

In this case $\omega_{3}^{2}$ is independent of $l$ and therefore of $L$.

\begin{eqnarray}
4) \ \ \ \omega_{4}^{2}(l,\beta) = (\frac{(5 - \beta)(l + 1) - 2\sqrt{4(l-1)^{2} + l}}{1 + l})M^2_{\phi},
\label{sn49}
\end{eqnarray}
with $l$ satisfying (\ref{sn35}) or
\begin{eqnarray}
{\rm sn}^2(\frac{M_{\phi}L}{2\sqrt{l+1}},l) = \frac{3}{2l + 2 -
\sqrt{4(l-1)^2 + l}}.
\label{sn50}
\end{eqnarray}

Again, in this case only $l = 1$ $(L = \infty)$ is solution of (\ref{sn50}). So $\omega^{2}_{4}$ is an allowed eigenvalue with $l$ satisfying (\ref{sn35}).

\begin{eqnarray}
5) \ \ \ \omega_{5}^{2}(l,\beta) = (\frac{2l + 5 -2\sqrt{l^{2} - l + 4}}{1 + l} - \beta)M^2_{\phi},
\label{sn51}
\end{eqnarray}
with $l$ satisfying an equation similar to (\ref{sn36}) or 
\begin{eqnarray}
{\rm sn}^2(\frac{M_{\phi}L}{2\sqrt{l+1}},l) = \frac{1}{l + 2 -
\sqrt{l^2 - l + 4}}.
\label{sn52}
\end{eqnarray}

As in the previous case, only $l = 1$ $(L = \infty)$ is solution of (\ref{sn52}).

\begin{eqnarray}
6) \ \ \ \omega_{6}^{2}(l,\beta) = (\frac{5l + 2 - 2\sqrt{4\,l^{2} - l + 1}}{1 + l} - \beta)M^2_{\phi},
\label{sn53}
\end{eqnarray}
with $l$ satisfying 
\begin{eqnarray}
{\rm sn}^2(\frac{M_{\phi}L}{2\sqrt{l + 1}},l) = \frac{1}{l} \ \ \ \ \ \ \ \
\mbox{or} \ \ \ \ \ \ \ \ {\rm sn}^2(\frac{M_{\phi}L}{2\sqrt{l+1}},l) = \frac{1}{2l + 1 - \sqrt{4l^2 - l + 1}}.
\label{sn54}
\end{eqnarray}

These equations are also satisfied only for $l = 1$ ($L = \infty $).

\begin{eqnarray}
7) \ \ \ \omega_{7}^{2}(l,\beta) = (\frac{(5 - \beta)(l + 1) + 2\sqrt{4(l-1)^{2} + l}}{1 + l})M^2_{\phi},
\label{sn55}
\end{eqnarray}
with $l$ satisfying (\ref{sn35}) or
\begin{eqnarray}
{\rm sn}^2(\frac{M_{\phi}L}{2\sqrt{l+1}},l) = \frac{3}{2l + 2 +
\sqrt{4(l-1)^2 + l}}.
\label{sn56}
\end{eqnarray}

By numerical analysis is possible to show that $M_{\phi}L$ belongs to the interval $(2.04,2.92)$. Thus these solutions must be discarded since $M_{\phi}L < 2\pi$. So $\omega^{2}_{7}$ is an allowed eigenvalue with $l$ satisfying (\ref{sn35}).

The above study shows that only the eigenvalues $\omega_{1}^{2}$ and $\omega_{2}^{2}$ are allowed for $n=2$, $\omega_{4}^{2}$ and $\omega_{7}^{2}$ for $n=3$. Also there is a trivial one, namely $\omega_{1}^{2}$ for $n=1$ which coincides with $\omega_{3}^{2}$ for $n=2$.

In the next section, we study the behavior of the energy eigenvalues $\omega^{2}$, for a fixed $\beta$, running continuously the external parameter of the theory $l \equiv l(L)$. 

\vspace*{0.9cm}

\noindent
{\large \bf 5. Level Shifts Induced by Box Size Changing and Points of \\ \hspace*{0.7cm}Instability}

\vspace*{0.5cm}

In this section we study the behavior of level shifts under changing of the box size. The case $n=1$, although has an allowed level, namely $\omega_{1}^{2}$, its does not depends on $L$ in a non-trivial way. So we do not consider it interesting to our study, in this work.

For the cases $n = 2$ and $3$ we have non-trivial results. In all cases below we fixed different values for the mass parameter $\beta$. Also we demand classical stability for the eigenfunctions $\psi_{i}$ $( i = 1,2,3)$. Classical stability means that the energy eigenvalues $\omega_{i}^{2}$ are non-negative [3], so that the amplitude of field $\chi$ does not grow exponentially in time. 
 
\vspace*{0.5cm}

\noindent
{\bf A) \ \ CASE \ $n = 2$ \ ($g = \frac{3}{2}\lambda$)}

\vspace*{0.3cm}

Considering only the energy eigenvalues  $\omega_{i}^{2} \geq 0$ in Eqs. (\ref{sn34}), (\ref{sn37}) and (\ref{sn39}) we obtain the following relations:

\vspace*{0.3cm}

\noindent
(a) \ \ $\beta \leq \frac{4 + l}{1 + l}$ \ \ \ \ \ for \ \ \ $\omega_{1}^{2}$,

\vspace*{0.2cm}

\noindent
(b) \ \ $\beta \leq \frac{4l + 1}{1 + l}$ \ \ \ \ for \ \ \ $\omega_{2}^{2}$,

\vspace*{0.2cm}

\noindent
(c) \ \ $\beta \leq 1$ \ \ \ \ \ \ \ \ for \ \ \ $\omega_{3}^{2}$.

\vspace*{0.3cm}

Since $l \in (0,1]$, from these relations we get the allowed intervals for  $\beta$. They are: 

\vspace*{0.3cm}

\noindent
(a) \ For \ $\omega_1$, \ $\beta \in [\frac{5}{2},4)$.

\vspace*{0.2cm}

\noindent
(b) \ For \ $\omega_2$, \ $\beta \in (1,\frac{5}{2}]$.

\vspace*{0.2cm}

\noindent
(c) \ For \ $\omega_3$, \ $\beta \in [0,1]$.

\newpage

Below we study the behavior of the energy eigenvalues $\omega^{2}_{i}$ under the 
running of the external parameter $L$. In order to do this we fix some particular values of the mass ratio parameter $\beta$.  

\vspace*{0.6cm}
 
\begin{figure}[ht]
\centerline{\psfig{figure=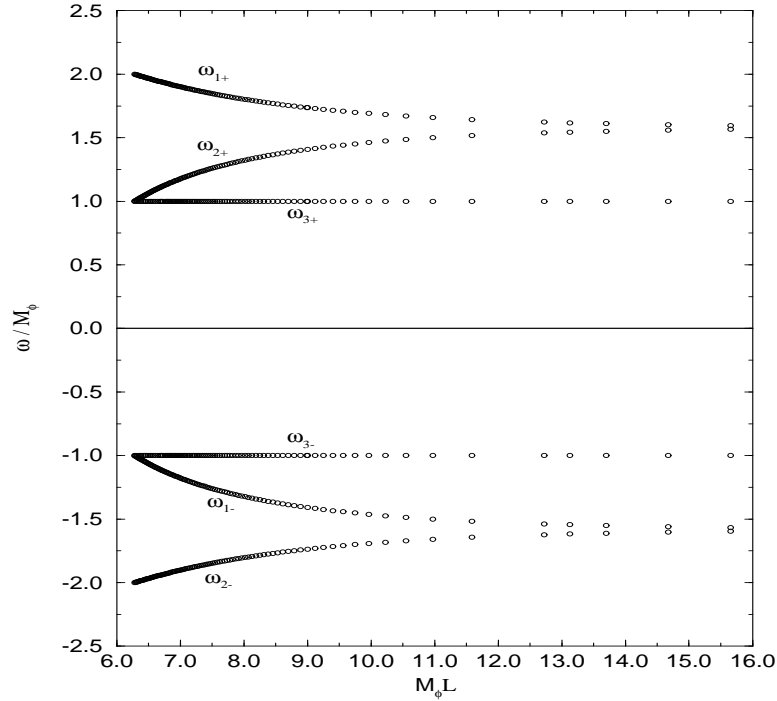,height=3.2in,width=3.2in}}
\vspace*{0.2cm}
\caption{The energy eigenvalues for $\beta =0$.}
\label{fig1}
\end{figure}

\noindent
(1) \ {\bf $\beta = 0$ \ (Fig. \ref{fig1})}

\vspace*{0.3cm}

\noindent
Using (\ref{sn34}), (\ref{sn37}) and (\ref{sn39}) we obtain

\vspace*{0.3cm}

\noindent
(a) \ $\omega^{2}_{1} = (\frac{4 + l}{1 + l})M^{2}_{\phi}$,

\vspace*{0.2cm}

\noindent
(b) \ $\omega^{2}_{2} = (\frac{4l + 1}{1 + l})M^{2}_{\phi}$,

\vspace*{0.2cm}

\noindent
(c) \ $\omega^{2}_{3} = M^{2}_{\phi},
\ \ \ \ \ \ \mbox{satisfied for all} \ L$.

\vspace*{0.3cm}

From the Fig. \ref{fig1} we can see that for large box $(l = 1 \ \mbox{or} \  
L=\infty)$, $\omega_1$ and $\omega_2$ coincide at 
$\omega^{2} = \frac{5}{2}M_{\phi}^{2}$. 

\vspace*{0.7cm}

\vspace*{0.5cm}

\begin{figure}[ht]
\centerline{\psfig{figure=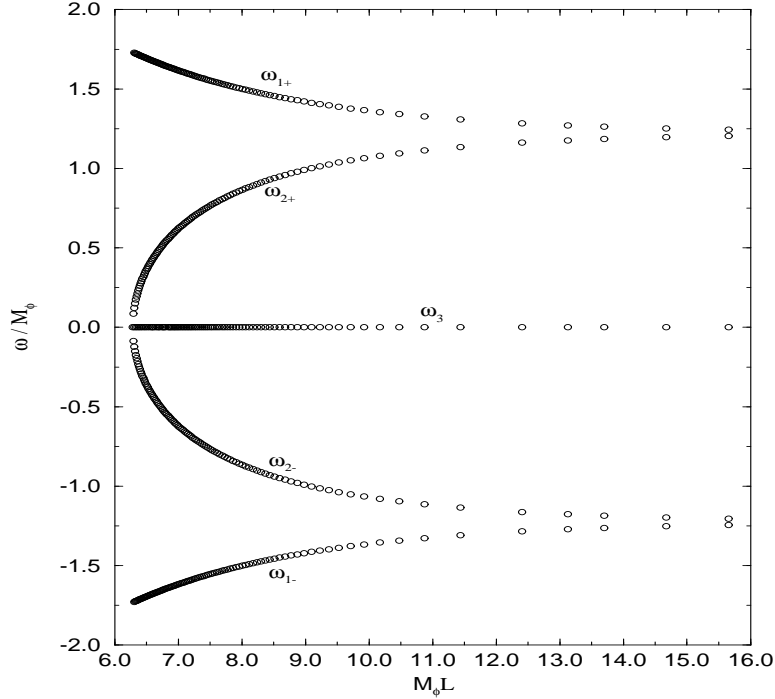,height=3.2in,width=3.2in}}
\vspace*{0.3cm}
\caption{The energy eigenvalues for $\beta =1$.}
\label{fig2}
\end{figure}

\noindent
(2) \ {\bf $\beta = 1$ \ (Fig. \ref{fig2})}

\vspace*{0.3cm}

\noindent
As before, from (\ref{sn34}), (\ref{sn37}) and (\ref{sn39}) we obtain

\vspace*{0.3cm}

\noindent
(a) \ $\omega^{2}_{1} = (\frac{3}{1 + l})M^{2}_{\phi}$,

\vspace*{0.2cm}

\noindent
(b) \ $\omega^{2}_{2} = (\frac{3l}{1 + l})M^{2}_{\phi}$,

\vspace*{0.2cm}

\noindent
(c) \ $\omega^{2}_{3} = 0$,
\ \ \ \ \ \ \mbox{satisfied for all L}.


\newpage

It is interesting to note that (see Fig. \ref{fig2}) for $L=\infty$, $\omega_{1}$ and $\omega_{2}$ converge to $\sqrt{\frac{3}{2}}M_{\phi}$, i.e., for large box $(L=\infty)$ the excited state of Dashen-Hasslacher-Neveu (DHN) [5] is obtained. Likewise the ground state of (DHN) model also is obtained, i.e., $\omega_{3} = 0$. This can be proved directly from Eq. (\ref{sn5}) taking $l = 1$. Nevertheless the Eq. (\ref{sn5}), or its equivalent Eq. (\ref{sn11}), has an aditional freedom by varying the parameter $\beta$.  

The Fig. \ref{fig2} shows that for $l \rightarrow 0$ 
($M_{\phi}L \rightarrow 2\pi$) [6], the energy eigenvalues $\omega_{2+}$ and 
$\omega_{2-}$ go to $\omega = 0$ for a critical size of the box, namely 
$L = \frac{2\pi}{M_{\phi}}$. This could suggest that this is an instability point of the field $\chi$, induced by changing the external parameter $L$ (box size). This is not the case here, because $l \rightarrow 0$ implies by Eqs. (\ref{sn3}) and (\ref{sn4}) that $\phi = 0$ and we are left only with a free Lagrangian of the $\chi$ field and no critical point exists. 

\vspace*{0.7cm}

\vspace*{0.5cm}

\begin{figure}[ht]
\centerline{\psfig{figure=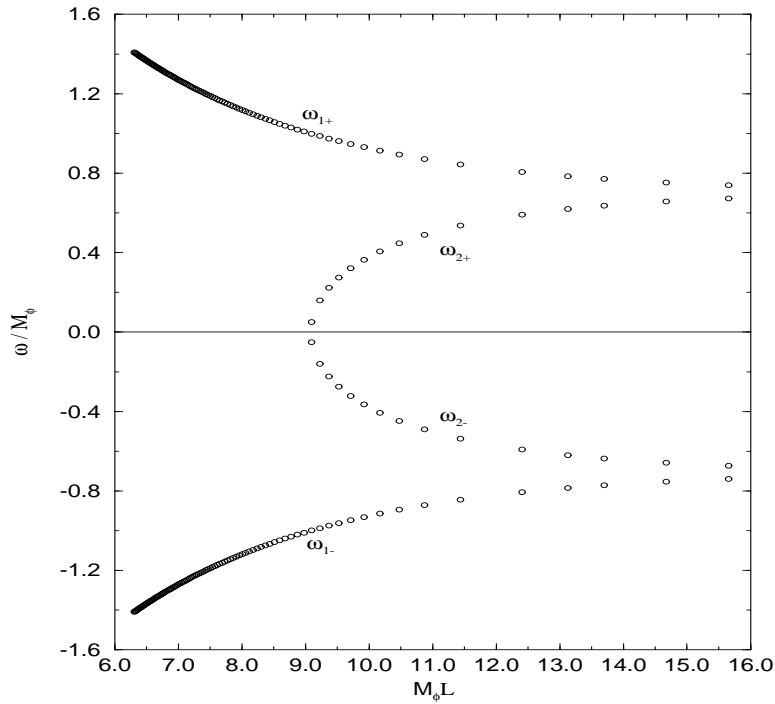,height=3.2in,width=3.2in}}
\vspace*{0.3cm}
\caption{The energy eigenvalues for $\beta =2$.}
\label{fig3}
\end{figure}

\noindent
(3) \ {\bf $\beta = 2$ \ (Fig. \ref{fig3})}

\vspace*{0.3cm}

\noindent
In this case, from (\ref{sn34}), (\ref{sn37}), (\ref{sn39}) we have

\vspace*{0.3cm}

\noindent
(a) \ $\omega^{2}_{1} = (\frac{2 - l}{1 + l})M^{2}_{\phi}$,

\vspace*{0.2cm}

\noindent
(b) \ $\omega^{2}_{2} = (\frac{2l - 1}{1 + l})M^{2}_{\phi}$,

\vspace*{0.2cm}

\noindent
(c) \ $\omega^{2}_{3} = -M^{2}_{\phi}$,
\ \ \ \ \ \ \mbox{satisfied for all L}. 


\newpage

Observe that $\omega_{3}^{2}$ is negative, so the classical configuration associated to this eigenvalue is unstable [3] and $\omega_{2+}$ turns out the new ground state. In the interval $l \in (0,\frac{1}{2})$ $\omega^{2}_{2}$ is negative. Thus its classical configuration is unstable in this interval. Also observe that now the instability point occurs for a bigger size (not the minimal one) of the box. Here ones can ask whether this kind of instability could lead to a ``condensate'' in a quantum regime. A full proof of this, of course,  requires a second quantization  for the field $\chi$, at least, in a semiclassical aproach. Also it is well known that inclusion of a non-linear term for it in the total Lagrangian could lead to quantum condensates [9]. These lines will not pursued here, since our aim is just to show the behavior energy levels for fields (in the classical limit) placed inside boxes, for a very simple geometry like an interval.  For $l = 1$ $(M_{\phi}L = \infty)$ $\omega_{1}$ and  $\omega_{2}$ coincide at  the value $\frac{\omega}{M_{\phi}} \sim 0.7$. See Fig. \ref{fig3} below. 

\vspace*{0.7cm}

\vspace*{0.5cm}

\begin{figure}[ht]
\centerline{\psfig{figure=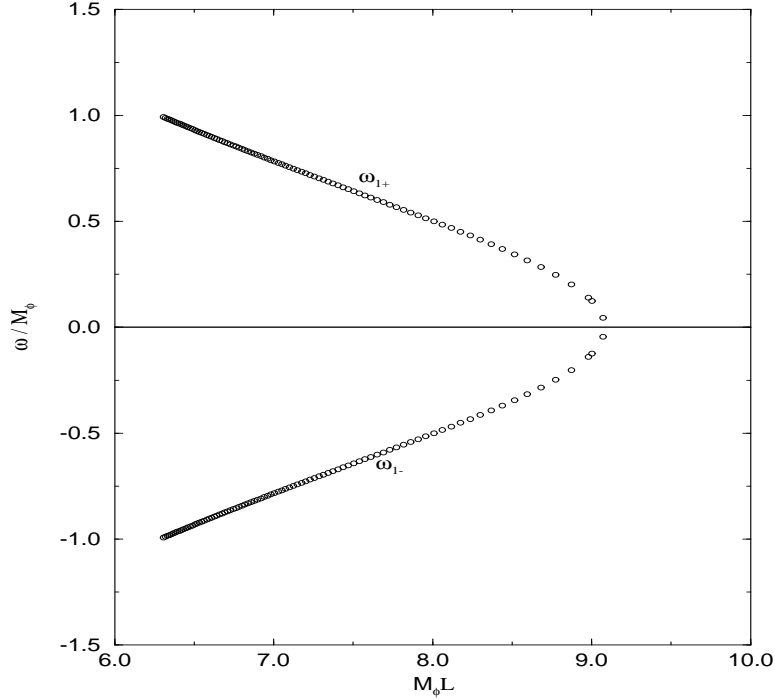,height=3.2in,width=3.2in}}
\vspace*{0.3cm}
\caption{The energy eigenvalues for $\beta =3$.}
\label{fig4}
\end{figure}

\noindent
(4) \ {\bf $\beta = 3$ \ (Fig. \ref{fig4})}

\vspace*{0.3cm}

\noindent
In this case, using (\ref{sn34}), (\ref{sn37}) and (\ref{sn39}) we have

\vspace*{0.3cm}

\noindent
(a) \ $\omega^{2}_{1} = (\frac{1 - 2l}{1 + l})M^{2}_{\phi}$,

\vspace*{0.2cm}

\noindent
(b) \ $\omega^{2}_{2} = (\frac{l - 2}{1 + l})M^{2}_{\phi}$,

\vspace*{0.2cm}

\noindent
(c) \ $\omega^{2}_{3} = -2M^{2}_{\phi}$,
\ \ \ \ \ \mbox{satisfied for all L}.


\newpage

As in previous cases, $\omega^{2}_{3}$ is negative. Besides, for 
$l \in(0,1]$ $\omega_{2}^{2} < 0$. So the classical configurations for 
$\omega_2$ and $\omega_3$ are unstable. Likewise in the interval 
$l \in(\frac{1}{2},1]$ we have that $\omega_{1}^{2} < 0$, so its classical 
configuration also is unstable in this interval. 
The Fig. \ref{fig4} shows that, the increasing of the mass parameter $\beta$ 
we get an instability point more and more near the critical size 
of the box $L = \frac{2\pi}{M_{\phi}}$. 

\vspace*{0.8cm}

\vspace*{0.8cm}

\noindent
(5) \ {\bf $\beta = 4$}

\vspace*{0.5cm}

\noindent
Using (\ref{sn34}), (\ref{sn37}) and (\ref{sn39}) we obtain

\vspace*{0.4cm}

\noindent
(a) \ $\omega^{2}_{1} = -(\frac{3\,l}{1 + l})M^{2}_{\phi}$,

\vspace*{0.2cm}

\noindent
(b) \ $\omega^{2}_{2} = -\frac{3}{1 + l}M^{2}_{\phi}$,

\vspace*{0.2cm}

\noindent
(c) \ $\omega^{2}_{3} = -3M^{2}_{\phi}$.

\vspace*{0.3cm}

We can see from these relations that all the energy eigenvalues are negative (for $l \neq 0$) and therefore their classical configurations will be unstable. 

It is interesting to note that for $\beta \in [2,3]$ the $\omega_2$ disappears for $\beta \rightarrow \frac{5}{2}$ and for $\frac{5}{2} < \beta < 3$ only $\omega_1$ survives and it is easy to see from Fig. \ref{fig4} above that its behavior is 
inverse of that one of $\omega_2$.

Therefore for $n=2$, a changing of the box size induces the appearing of an instability point for the energy eigenvalues $\omega_1$ or $\omega_2$ for $\beta \in (1,4)$. 

\vspace*{0.5cm}

\noindent
{\bf B) \ \ CASE \ $n = 3$ \ ($g = 3\lambda$)}

\vspace*{0.3cm}

For the case $n = 3$ the results are pretty  the same to its seven bound energy levels. However, in this case, only the eigenfunctions given by Eqs. (\ref{sn27}) and (\ref{sn30}) with respective  energy eigenvalues given by Eqs. (\ref{sn49}) and (\ref{sn55}) satisfy the condition (\ref{sn6}). The only great difference is that there is no instability point for the mass ratio parameter $\beta \in (4,6)$. In fact the instability points only exists for $\beta \in (1,4]$ or $\beta \in [6,9)$.

Another interesting point, as showed in [6], is that imposing periodic boundary conditions at $x = \pm\,\frac{L}{2}$ on the field $\phi$, the same relation (\ref{sn6}) is obtained and therefore all results from the  Dirichlet's case keep valid. Also imposing periodic boundary conditions at $x = \pm\,\frac{L}{2}$ on the solutions $\psi_{s}$, it is possible to show that for the case $n = 2$, the eigenfunctions given by Eqs. (\ref{sn20}) and (\ref{sn21}) lead to the same relation for $l \equiv l(L)$ given by (\ref{sn6}). In same way, for the case $n = 3$, the eigenfunctions given by Eqs. (\ref{sn27}) and (\ref{sn30})  also lead to (\ref{sn6}). Therefore all results of the Dirichlet case keep valid for periodic boundary conditions.

\vspace*{0.9cm}

\noindent
{\large \bf 6. Conclusions}

\vspace*{0.5cm}

In this work we studied the energy eigenvalues $\omega^{2}$ of a classical scalar field $\chi$ in $(1 + 1)$ dimension  interacting with another classical scalar field $\phi$ through the Lagrangian ${\cal L}_{int} = g\phi^{2}\chi^{2}$, in a finite domain (box of size $L$). The energy eigenvalues depend on four parameters, namely, $\beta$ (mass ratio parameter), coupling constants $\lambda$ and $g$, and $l$ (which is connected with the box size $L$). We fixed the coupling constant $g$ by Eq. (\ref{sn10}) for an arbitrary $\lambda$ and we studied only the cases $n =1,2,3$, which correspond to a moderate strengh interaction constant $g$ related to $\lambda$. For the more general case of $n$ real a full numerical treatment perhaps is necessary. Next, we discussed the behavior of the energy eigenvalues $\omega^{2}$ by fixing the parameter $\beta$ and changing the external parameter of the theory $l \equiv l(L)$, namely the size of the box.    

In the case $n = 2$ $(g = \frac{\lambda}{2})$, we concluded that the instability points for the energy eigenvalues $\omega_1$ or $\omega_2$ occur for $\beta \in (1,4)$. These instability points are obtained as a consequence of squeezing the box. Also, in Figs. \ref{fig2}, \ref{fig3} and \ref{fig4} it is shown that by varying the mass ratio $\beta$, while $\omega_1$ presents its instability point more and more near the minimal size of the box $L = \frac{2\pi}{M_{\phi}}$, $\omega_2$ presents its one more and more far from this value.

For the case $n = 3$ $(g = 3\lambda)$, we have seven bound energy levels. Also only two of them satisfy the Dirichlet's boundary  condition which in turn implies Eq. (\ref{sn6}). Their behavior are pretty the same of $n = 2$, but the instability points only exists for $\beta \in (1,4]$ or $\beta \in [6,9)$. For $\beta$ in interval $(4,6)$ we get only stable solutions. For periodic boundary conditions all results obtained with Dirichlet's boundary conditions keep unchangeable. Of course other boundary conditions can be imposed leading to new behaviors of the energy levels under box squeezing.

Several interesting extensions and approaches can  be pursued from this work.
As stressed in the introduction, fields placed in cavities lead to new and sometimes unexpected behaviors of some systems. Although our approach is for classical fields it suggests that a quantization of the system above studied could lead to formation of a kind of ``condensate" just by squeezing the system in a box. Of course this would require at least a semiclassical approach which will be done elsewhere. A more interesting case would be the inclusion of non-linearities for the classical field $\chi$, which leads to well known condensates for unbound domains [9], but now in finite domains. Also in this case we think a full numerical treatment is needed. 

Generalization of the above results to $n$ spatial dimensions leads to more complex equations,  not to mention that we have, in this case, an enormous (infinite to be sure) variety of geometries for the shape of the box. Nevertheless these kind of calculations for spherical symmetry could be interesting in order to study for example bound states behavior of matter fields in compact stars and in reheating theory and in inflationary cosmology.

\vspace*{0.8cm}

\noindent
{\large \bf Acknowledgments}

\vspace*{0.5cm}

This work was supported, in part, by FAPESP (Funda\c{c}\~ao de 
Amparo \`a Pesquisa do Estado de S\~ao Paulo), Brazil, under contract 97/04248-2. The authors are grateful to Professor A. Grib for his kind suggestions. 

\vspace*{0.8cm}

\noindent
{\large \bf Appendix}

\vspace*{0.5cm}

\hspace*{0.6cm}We show below a sufficient condition for that the associate Hamiltonian to the Lagrangian (\ref{p}) be positive definite. 

\hspace*{0.6cm}We consider the potencial for the field $\chi$, given by:
\begin{eqnarray*}
V(\chi) = - \frac{1}{2}M_{\chi}^{2}\chi^2 + g\phi^2\chi^2,
\end{eqnarray*}
we have
\begin{eqnarray*}
V^{\prime}(\chi) = -M_{\chi}^{2}\chi + 2g\phi^2\chi = 0 \ \ \Rightarrow \ \ \chi = 0 \ \ \mbox{(critical point).}  
\end{eqnarray*}

The second derivate of the potencial $V(\chi)$ is given by
\begin{eqnarray*}
V^{\prime\prime}(\chi) = -M_{\chi}^{2} + 2g\phi^2.
\end{eqnarray*}

In order to $V^{\prime\prime}(\chi) > 0$ we must impose that
\begin{eqnarray}
\phi^2 > \frac{M_{\chi}^{2}}{2g}.
\label{Q}
\end{eqnarray}

Thus the above condition leads to the existence of a state of least energy (vacuum) of the field $\chi$.

On the other hand, the Hamiltonian for the field $\chi$ is given by
\begin{eqnarray*}
H = \dot{{\chi}}^{2} - {\cal{L}}_{\chi} = \dot{\chi}^{2} - \frac{1}{2}\partial_{\mu}\chi\partial^{\mu}\chi - \frac{1}{2}M_{\chi}^{2}\chi^2 + g\phi^{2}\chi^{2},
\end{eqnarray*}
from there we have that
\begin{eqnarray*}
H = \frac{1}{2}\dot{\chi}^{2} \ + \ \frac{1}{2}\frac{d^{2}\chi}{dx^{2}} \ + \ (g\phi^2 \ - \ \frac{1}{2}M_{\chi}^{2})\chi^2,
\end{eqnarray*}
which will be positive definite, if the condition (\ref{Q}) is satisfied.


\vspace*{0.8cm}

\noindent
{\large \bf References}

\vspace*{0.5cm}

\noindent 
[1] G. Plunien, B. M\"uller and W. Greiner, Phys. Rep. ${\bf 134}$, 87 (1986).\newline
[2] S. Haroche and D. Kleppner, Phys. Today ${\bf 42}$, 24 (1989).\newline
[3] R. Jackiw, Rev. Mod. Phys. ${\bf 49}$, 681 (1977).\newline
[4] J. A. Espich\'an Carrillo and A. Maia Jr., ``Energy Levels of Interacting Fields in a Box", Int. \hspace*{0.6cm}J. Theor. Phys., Vol. {\bf 38}, $N^{0}$ 8, 2183 (1999), hep-th/9905158. \newline
[5] R. Dashen, B. Hasslacher and A. Neveu, Phys. Rev. D {\bf 10}, 4131 (1974).\newline
[6] J. A. Espich\'an Carrillo, A. Maia Jr. and V. M. Mostepanenko, ``Jacobi
Elliptic Solutions of \hspace*{0.6cm}$\lambda \phi^{4}$ Theory in a Finite Domain", accepted for publication in Int. J. Mod. Phys. A (1999), 
\hspace*{0.6cm}hep-th/9905151. \newline
[7] M. Abramowitz and I. A. Stegun, Hanbook of Mathematical Functions, Dover Publications, \hspace*{0.6cm}INC., New York, (1972).\newline
[8] Z. X. Wang and D. R. Guo, Special Functions, World Scientific (1989).\newline
[9] A. A. Grib, S. G. Mamayev and V. M. Mostepanenko, Vacuum Quantum Effects in Strong \hspace*{0.6cm}Fields, Friedmann Laboratory Publishing, St. Petersburg, (1994).


 \end{document}